\documentstyle[preprint,aps,prd,epsf]{revtex}
\tighten

\begin{document}
\draft
\title{Does the quark-gluon plasma contain stable hadronic bubbles?}
\author{Gregers Neergaard and Jes Madsen}
\address{Institute of Physics and Astronomy, University of Aarhus,
DK-8000 {\AA}rhus C, Denmark}

\date{\today}

\maketitle

\begin{abstract}
We calculate the thermodynamic potential of 
bubbles of hadrons embedded in quark-gluon plasma, and of 
droplets of quark-gluon plasma embedded in hadron phase.
This is a generalization of our previous results 
to the case of non-zero chemical potentials.
As in the zero chemical potential case, we find that
a quark-gluon plasma in thermodynamic equilibrium may
contain stable bubbles of hadrons of radius $R \simeq 1$~fm.
The calculations are performed within the MIT Bag model,
using an improved multiple reflection expansion.
The results are of relevance for neutron star phenomenology
and for ultrarelativistic heavy ion collisions.
\end{abstract}



\section{Introduction}
\label{sec.intro}
In a previous paper \cite{us} we calculated the free energy
of a bubble of hadrons embedded in an extended quark-gluon plasma (QGP),
and of a droplet of QGP embedded in an extended hadron phase, 
for parameters in the vicinity 
of the cosmological quark-hadron transition,
i.e.\ the baryon chemical potential
were set to zero.
In Ref. \cite{us} we found, as Mardor and Svetitsky in \cite{marsve}, 
that the free energy of a hadron bubble of radius $R$ 
embedded in QGP possessed a minimum at radii of a few fm,
even above the phase transition temperature.
Thus, hadronization from QGP is strongly enhanced compared to the 
usual nucleation scenario, where an energy barrier has to be overcome
before bubbles of hadrons can form and grow.

An important ingredient in these calculations is the density of states
of the relevant particles. In Ref.\ \cite{us} we advised 
a modification of the usual expressions arising from the
multiple reflection expansion (MRE), and we saw that
this modification (the MMRE) yielded more accurately
the free energy than the MRE did when compared to a direct (but numerically 
heavier) sum-over-discrete-states calculation.
Our modification of the multiple reflection expansion expressions
is physically well motivated, in that it 
consists 
of a truncation of the density of states 
in such a way that we avoid the effects of a negative density
of states, present in the usual MRE expressions.

In this paper, we generalize the calculations to finite chemical
potential, and we shall see that also in the case of non-zero chemical 
potential, the MMRE produces accurately the thermodynamic potential.
We show that, as in the case of zero chemical potential,
the thermodynamic potential of a hadron bubble embedded in QGP 
has a minimum at a radius of $R \simeq 1$~fm, meaning that
it is thermodynamically favorable for a QGP in equilibrium 
to spontaneously create bubbles of hadrons of about this size. 
However, the minimum in the thermodynamic potential
found for non-zero chemical potential is less pronounced
than in the zero chemical potential case considered in Ref.~\cite{us}.
The more general calculations presented in this paper 
should be relevant also to neutron star phenomenology
as well as  ultrarelativistic heavy ion collisions.

In the following section we give an overview of the basic theory
leading to the results of section~\ref{sec.results}.
Finally in Sec.~\ref{sec.concl}, we summarize the conclusions.

\section{Theoretical framework}
\label{sec.theory}
\subsection{The particle model}
\subsubsection{The quark-gluon plasma}
The model of the plasma phase is a close derivative
of the MIT Bag model \cite{bagmodel1,bagmodel2}.
It consists essentially of non-interacting quarks and gluons
(also self-interaction of the gluons is neglected),
bounded by MIT Bag-like boundary conditions\cite{csc-comment}.
There are three quark flavors in our model: 
$u$ and $d$, which we consider massless, and
$s$, to which we assign a mass of $m_s=150$~MeV.
The other quarks are too heavy to be relevant in this analysis.
We choose a critical temperature (the bulk phase equilibrium temperature
at zero chemical potential) of $T=150$~MeV, which gives a 
Bag constant of $B=(221$~MeV$)^4$.
The plasma may be inside a sphere, as in the usual MIT Bag
(we shall refer to this configuration as a plasma droplet),
or outside, i.e.\ in a ``vacuum bubble'' configuration. 
For a detailed review of this model, the reader may consult \cite{us}.
We are mostly interested in the vacuum bubble situation, 
where the vacuum bubble is filled
with a thermal mixture of hadrons as described later.

For the density of states of the quarks and gluons we use 
a modification of the multiple reflection expansion (MRE)
\cite{bb1,berger,farhi,madsen,bb2}, 
advised in~\cite{us}.
This modification affects only the gluon density of states,
which is then 
\begin{equation}
\rho_g(k,R) = 
  \left\{ \begin{array}{ll}0, & 0 \leq kR<0.832 \\
  \frac{V_{QGP}k^2}{2\pi^2}-\frac{4R}{3\pi}, & kR \geq 0.832
\end{array}
\right.
\label{rhog}
\end{equation}
for gluons inside a sphere (i.e.\ the MMRE($R$)), and
\begin{equation}
\rho_g(k,R) = 
  \left\{ \begin{array}{ll}0, & 0 \leq kR<0.458 \\
  \frac{-V_H k^2}{2\pi^2}+\frac{4R}{3\pi}, & kR \geq 0.458
\end{array}
\right.
\label{rhog-r}
\end{equation}
for the difference between the gluon density of states
in a volume $V_\infty$ and in volume $V_\infty-V_H$,
(i.e.\ the MMRE($-R$)).

\begin{figure}
\epsfxsize=8.6truecm
\epsfbox{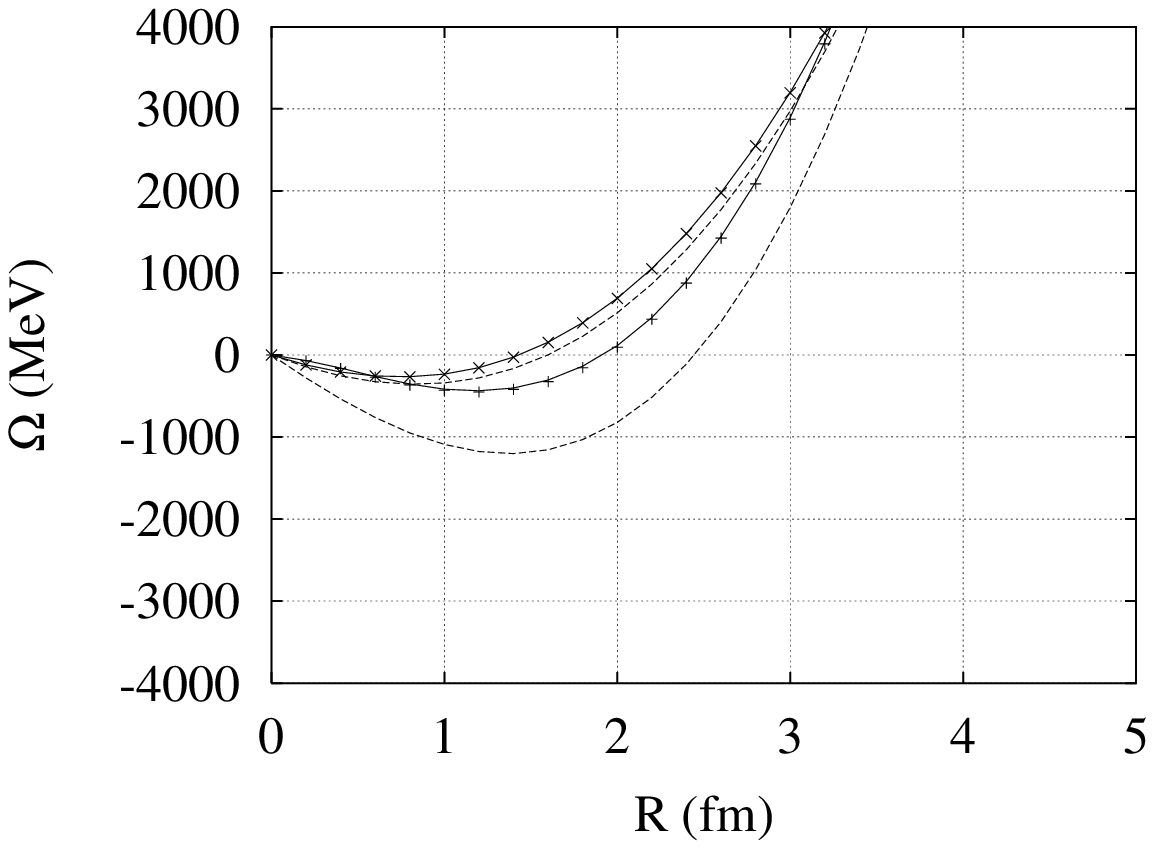}
\caption[]{
Thermodynamic potential of a bubble of hadron phase of radius $R$ embedded
in an extended QGP phase. 
In the two upper curves $(\mu_q,T)=(450,46)$~MeV, i.e.\ a 
higher temperature than the phase transition temperature
at this chemical potential, $T_0=41.0$~MeV. 
In the two lower curves $(\mu_q,T)=(200,139)$~MeV, again slightly
above the phase transition temperature $T_0=133.7$~MeV.
To draw the solid lines, the MMRE($-R$) has been used to describe
the density of states in the expression for the thermodynamic potential,
whereas the dashed lines are for the MRE($-R$).
Based on comparisons with more direct
sum-over-states like calculations (the points practically coinciding
with the MMRE($-R$) lines),
we conclude that the MMRE($-R$) is the more correct
model for the density of states.}
\label{fig:mrecomp}
\end{figure}

In~\cite{us} we saw that this modified multiple reflection expansion (MMRE)
description of the density of states
made the free energy agree nicely with more direct calculations,
i.e.\  summing over discrete energy levels.
Before using the MMRE in the more general case under consideration
in this paper, we performed similar checks of the MMRE against direct
sum-over-discrete-states calculations. 
Also in this case, the MMRE proved to describe the density of states
adequately and significantly better than the MRE, cf.\ Fig.~\ref{fig:mrecomp}.
For a complete description of the MMRE density of states, 
we refer the reader to \cite{us}.

\subsubsection{The hadron phase}
We include all hadrons with masses below $1.2$~GeV,
taking only the volume part 
of the density of states into account.
This is justified as the hadrons represent far fewer degrees
of freedom than the QGP. 
Specifically, the density of states of a hadron species occupying
a volume $V_H=\frac{4\pi}{3}R^3$ is
\begin{equation}
\rho_H(k,R) = \frac{V_Hk^2}{2\pi^2}.
\label{rhoh}
\end{equation}

\subsection{Thermodynamics}
The thermodynamic potential is
\begin{equation}
\Omega(T,V,\{\mu_i\}) = -T \ln(Z(T,V,\{\mu_i\})),
\label{omega}
\end{equation}
where $Z(T,V,\{\mu_i\})$ is the partition function
\begin{equation}
Z(T,V,\{\mu_i\}) = 
Tr \left\{ e^{-({\mathcal H}-\sum_i \mu_i \Lambda_i)/T} \right\}.
\label{Z}
\end{equation}
Here, ${\mathcal H}$ is the Hamiltonian, 
the $\Lambda_i$'s are the symmetries 
of the Hamiltonian (the conserved quantities), and 
$\{\mu_i\}$ denotes the corresponding collection of chemical potentials.
$Tr\{.. \}$ means the sum over all diagonal elements of 
energy- and number-eigenstates. These are simultaneous eigenstates since 
$[{\mathcal H},\Lambda_i]=0$. The volume dependence enters via the 
energy levels, i.e.\ in the trace operation.

In thermodynamic equilibrium, the configuration realized in Nature
is the one which minimizes the thermodynamic potential,
leading to the Gibbs conditions for phase equilibrium: 
$T_1=T_2$ (thermal equilibrium),
$-\frac{\partial \Omega_1}{\partial V_1} = -\frac{\partial \Omega_2}{\partial V_2}$ (mechanical equilibrium), and 
$\mu_i^{(1)}=\mu_i^{(2)}$ (chemical equilibrium), 
where index $1$ and $2$ refer to the two phases.

If there are no interactions between the particles, i.e.\ the energy
of a particular state is the sum of the energies of the individual
particles, then we can write the thermodynamic potential 
for a particular particle species enclosed in a spherical volume
of radius $R$, as
\begin{equation}
\Omega(R,T,\mu) = \mp g T \int_0^{\infty} dk \, 
\rho(k,R) \ln(1 \pm e^{-(\sqrt{k^2+m^2}-\mu)/T}),
\label{omega.fse}
\end{equation}
(upper sign for fermions, lower sign for bosons)
where we further assume that the energy levels are sufficiently closely
spaced that we may use a smoothed density of states, $\rho(k,R)$, 
(e.g.\ the MMRE) normalized such that 
$\int_0^\infty dk\,\rho(k,R)$
is the total number of states in the volume $V=\frac{4\pi}{3}R^3$.
The chemical potential entering in~(\ref{omega.fse}), is the combined
chemical potential of that particle species
\begin{equation}
\mu = \sum_i \lambda_i \mu_i,
\label{mutot}
\end{equation}
$\lambda_i$ being the quantum expectation value of the
symmetry operator $\Lambda_i$ of the particle species in question 
(e.g.\ $\lambda_{\mathrm baryon}=1/3$ for a quark).
Finally, $g$ is an appropriate degeneracy factor.

At the energies relevant in this analysis, the quarks and gluons
can be considered point-like. This is not the case for the hadrons.
We must take explicit account of the fact that the hadrons occupy
a certain volume.
A number of such excluded volume corrections have been discussed
in the literature, but
we shall not go into an analysis of the different suggestions.
We choose the one proposed in \cite{hagraf},
since it is very easy to implement and grasps much of the essential
physics involved. According to \cite{hagraf}, 
the ``true'' pressure of the hadron gas ($p_H$) is obtained
from the ideal gas pressure ($p_{H,id}$) according to
\begin{equation}
p_H = \frac{p_{H,id}}{1+\epsilon_{H,id}/(4B)},
\label{pcorr}
\end{equation}
where 
\begin{equation}
\epsilon_{H,id} = \sum_i \epsilon_{i,id}
\label{esum}
\end{equation}
is the total energy density of the ideal hadron gas,
\begin{equation}
p_{H,id} = \sum_i p_{i,id}
\label{psum}
\end{equation}
is the total hadronic ideal gas pressure, 
($i \in {\mathrm hadron\ species}$),
and $B$ is the Bag constant. 
We note that some sort of excluded volume correction is essential,
since in a point-like hadron gas, no transition
to QGP will occur at low temperatures, even at arbitrarily high density.

There is one more question about the 
hadrons we need to address before we proceed.
This is the phenomenon of Bose-Einstein condensation,
regarding the bosonic part of the hadron gas.
A bosonic gas always has $\mu_i \leq m_i$, and when $\mu_i=m_i$
particles will start to ``condense'' into the lowest, zero-momentum
state, meaning that there will be a macroscopic number of
particles in this state. 
When the parameters are such that we would have $\mu_i \geq m_i$,
the hadron species $i$ 
is removed from the above summations~(\ref{esum}) and (\ref{psum}),
since particles in a zero-momentum state do not contribute any pressure.

\section{Results}
\label{sec.results}
We shall consider both the case of equal chemical potentials, 
$\mu_u=\mu_d=\mu_s \equiv \mu_q$,
and the case where the $u$ and $d$ quarks have a different chemical
potential from that of the $s$ quark, $\mu_u=\mu_d \neq \mu_s$.
The first case is relevant when
the time scales are such that the weak interactions maintain
equilibrium between all three flavors of quarks (e.g.\ quark matter
in neutron stars, 
neglecting the electron chemical potential as a first approximation), 
whereas the case of separate $s$ quark chemical potential is relevant
to ultrarelativistic heavy ion collisions, the time scales here being 
such that the net number of $s$ quarks is conserved separately
from that of the net number of light quarks.
In our model, the two light quarks are both massless, and since
we neglect electromagnetic interactions, there is no
difference between $u$ and $d$ quarks.

\subsection{Quark-hadron phase equilibrium}
\label{sec.pheq}
\begin{figure}
\epsfxsize=8.6truecm
\epsfbox{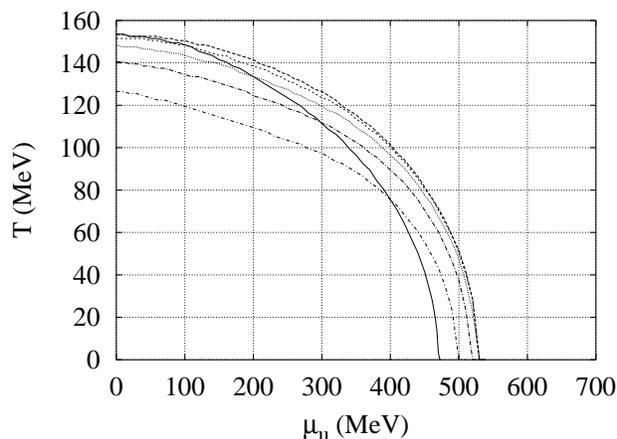}
\caption[]{
Phase equilibrium lines in the $(\mu_u,T)$-plane. 
At these lines, QGP and hadrons are in thermodynamic equilibrium.
The solid line represents the case of equal
quark chemical potentials, $\mu_u=\mu_d=\mu_s$, whereas the other lines
represent different values of the $s$-quark chemical potential: 
Top to bottom, $\mu_s=0,100,200,300,400$~MeV.}
\label{fig:pheq2}
\end{figure}

To study the phase equilibrium between the hadrons and the QGP,
we implement the Gibbs conditions
on the volume (or bulk) part of the thermodynamic potential,
i.e.\ also in the QGP density of states we only use the 
terms proportional to the volume.
(We shall see later that the surface terms give rise to
interesting features near bulk phase equilibrium.)
In this way, we obtain the phase diagram in Fig.~\ref{fig:pheq2}.
The QGP phase (above the phase equilibrium lines) 
occupies a larger and larger region of the phase diagram as 
the $s$-quark chemical potential is raised.
As one might have expected, 
the phase equilibrium line in the case of equal chemical potentials 
lies somewhere between the lines of varying $\mu_s$.

\subsection{The thermodynamic potential}
\label{sec.thermpot}
\begin{figure}
\epsfxsize=8.6truecm
\epsfbox{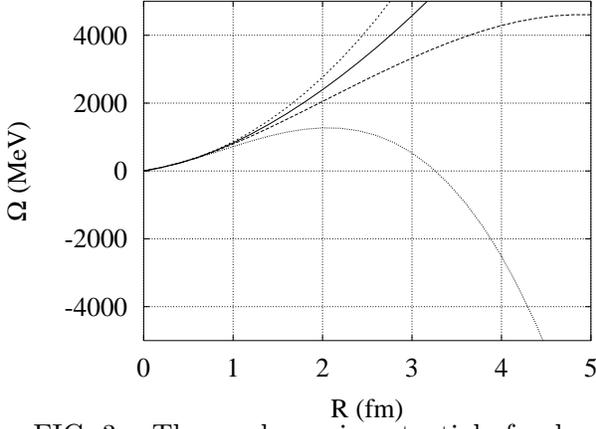}
\caption[]{
Thermodynamic potential of a droplet of QGP of radius $R$ embedded
in an extended phase of hadrons, normalized such that the thermodynamic 
potential of a pure hadron phase is zero. 
The plasma density of states is described by the MMRE($R$).
The bulk phase transition point is $(\mu_q,T)=(320,106.1)$~MeV
(cf.\ Fig.~\ref{fig:pheq2}). 
Curves are shown for different temperatures, top to bottom: 
$T=104,106.1,108,112$~MeV. 
The chemical potential is fixed at $\mu_q=320$~MeV.}
\label{fig:opla320}
\end{figure}

\begin{figure}
\epsfxsize=8.6truecm
\epsfbox{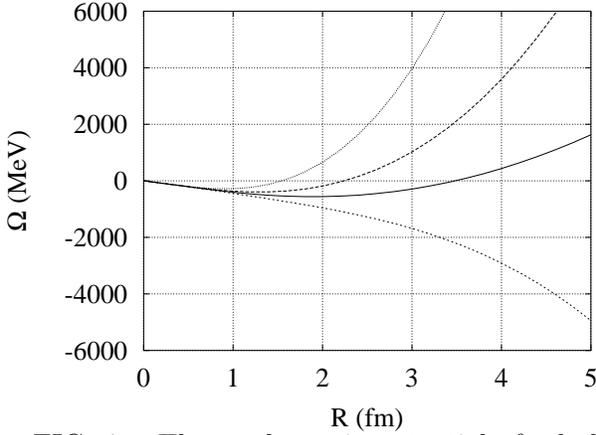}
\caption[]{
Thermodynamic potential of a bubble of hadron phase of radius $R$ embedded
in an extended QGP phase, normalized such that the thermodynamic 
potential of a pure QGP phase is zero. 
The plasma density of states is described by the MMRE($-R$).
The bulk phase transition point is $(\mu_q,T)=(320,106.1)$~MeV
(cf.\ Fig.~\ref{fig:pheq2}).
The chemical potential is fixed at $\mu_q=320$~MeV, and the temperature
varies, top to bottom: $T=112,108,106.1,104$~MeV.}
\label{fig:ovac320}
\end{figure}

\begin{figure}
\epsfxsize=8.6truecm
\epsfbox{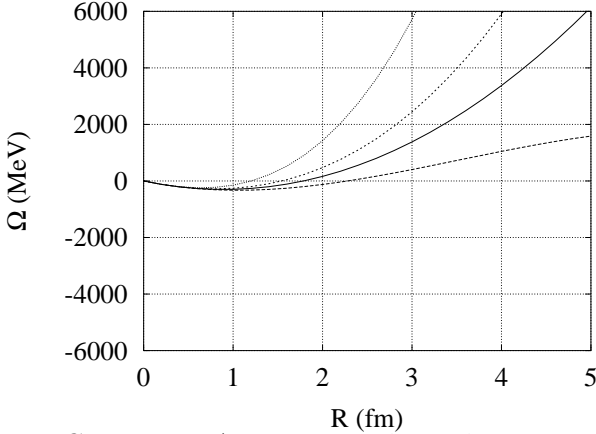}
\caption[]{
As Fig.~\ref{fig:ovac320}, but this time investigating the
bulk phase transition point $(\mu_q,T)=(450,41.0)$~MeV.
The chemical potential is fixed at this value, and
the temperatures are, top to bottom: $T=52,44,41.0,38$~MeV.}
\label{fig:ovac450}
\end{figure}

\begin{figure}
\epsfxsize=8.6truecm
\epsfbox{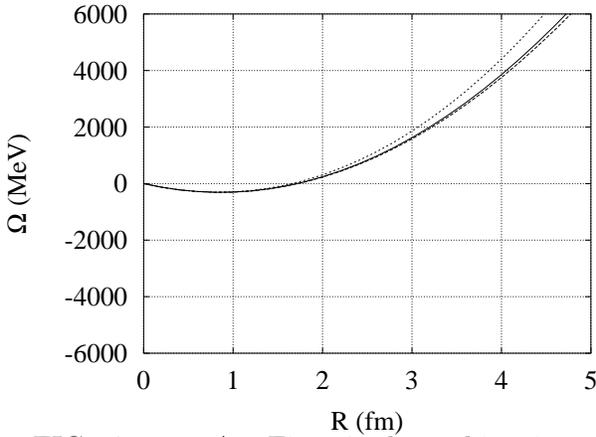}
\caption[]{
As Fig.~\ref{fig:ovac320}, but this time investigating the
bulk phase transition point $(\mu_q,T)=(470.0,3.28)$~MeV.
The chemical potential is fixed at this value, and
the temperatures are, top to bottom: $T=8,3.28,1$~MeV.
Notice that the depth of the minimum at $R\simeq 1$~fm is
practically independent of which equilibrium point we consider, 
cf.\ Figs.~\ref{fig:ovac320} and \ref{fig:ovac450}.
The curves in this figure differ somewhat in their qualitative behaviour
from those in Figs.~\ref{fig:ovac320} and \ref{fig:ovac450}.
This is because at these low temperatures, the surface terms
dominate over the volume terms at the radii shown here;
at larger radii the $T=1$~MeV curve bends over and eventually
tends to $-\infty$ as it should.}
\label{fig:ovac470}
\end{figure}

In this section we look at the thermodynamic potential 
in the vicinity of equilibrium, now using
the full expression of the density of states of the QGP
(i.e.\ including the surface terms).
In Figs.~\ref{fig:opla320}--\ref{fig:ovac470}, 
there is a single chemical potential, $\mu_q$,
common to all three quark flavors.

First we show Fig.~\ref{fig:opla320}, the thermodynamic potential
of a QGP droplet in equilibrium with an extended hadron phase near the
bulk equilibrium point $(\mu_q,T)=(320,106.1)$MeV.
The picture is as expected for a first order phase transition:
Below the bulk equilibrium temperature
no stable QGP droplet can form, and even somewhat above this temperature
there is an energy barrier for the system to pass 
before the true minimum at $R=\infty$ can be reached.
The surface terms are responsible for this energy barrier.
Precisely {\em at} the phase transition temperature the volume terms
cancel (by definition), and the thermodynamic potential goes to
infinity as $R^2$ (the leading surface term).

The more interesting conclusions are reached when one considers
the reverse situation, namely a hadron bubble in equilibrium
with an extended phase of QGP. 
Figs.~\ref{fig:ovac320}, \ref{fig:ovac450} and \ref{fig:ovac470}
show the thermodynamic potential of this configuration near 
different points on the phase equilibrium line.
In all three cases the potential exhibits a minimum at a radius of
$R \simeq 1$~fm of approximately the same depth, even at temperatures
above phase equilibrium temperature.
Such a minimum would not be present in the standard textbook treatment
of phase transitions, in which a first order phase transition usually
is described in terms of a phenomenological free energy containing
only volume- and (positive) surface terms.

Thus, a QGP in thermodynamic equilibrium apparently 
contains bubbles of hadrons, 
and the transition from QGP to hadrons will proceed 
(in addition to ordinary bubble nucleation and subsequent 
expansion of these bubbles)
by smooth expansion of the pre-formed bubbles.
In addition to this smooth growth as 
the minimum tends to larger radii when the temperature drops, 
pre-formed bubbles can also grow discontinuously by passing
the barrier from $R>0$ to some $R$ outside the barrier.
This possibility resembles very much the ordinary
nucleation scenario with a modified barrier height.
At the parameters of Figs.~\ref{fig:ovac320}, \ref{fig:ovac450}
and \ref{fig:ovac470}, the
mean number of quarks in a hadron bubble of radius $R=1$~fm
is typically 3--4, suggesting
that formation of a real hadron is not a very rare event.
However, the minimum being of modest depth, this interesting
effect could be due to the model's inaccurate representation of QCD.

The fact that there is no energy barrier for a hadron bubble to form in QGP 
does not imply that the phase transition is second order, or a smooth
cross-over. 
In the model adopted here, the transition is inherently first order,
since there is an {\em a priori} difference in entropy 
between the two bulk phases, and thus a latent heat.
As seen from e.g.\ Fig.~\ref{fig:ovac320} it is not energetically
favorable for bulk hadronic matter to form at temperatures above the
phase transition temperature $T_0$.
But the formation of isolated bubbles of hadronic matter {\em is} favorable
above $T_0$, so the transition will appear smoother than normally
expected for a first order transition, with a more gradual release of
latent heat.

\begin{figure}
\epsfxsize=8.6truecm
\epsfbox{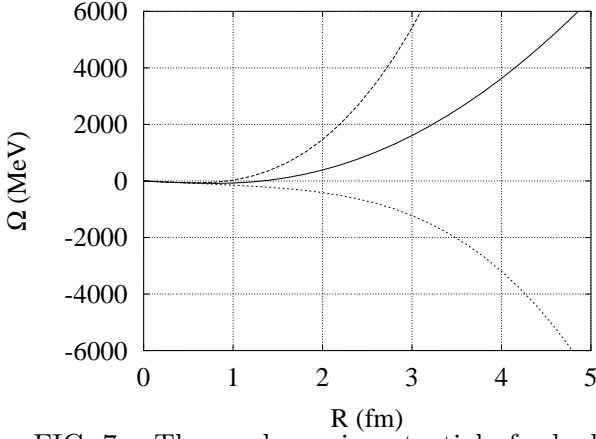}
\caption[]{
Thermodynamic potential of a hadron bubble embedded in an
extended QGP phase in the vicinity of the phase equilibrium
point $\mu_u=100$~MeV, $\mu_s=400$~MeV, $T_0=119.5$~MeV 
(cf.\ Fig.~\ref{fig:pheq2}). The chemical potentials are fixed
at these values, and the temperature varies, top to bottom:
$T=125,119.5,115$~MeV.
Notice that, at these values of the chemical potentials,
the minimum in the thermodynamic potential practically disappears.}
\label{fig:o2vac100}
\end{figure}

\begin{figure}
\epsfxsize=8.6truecm
\epsfbox{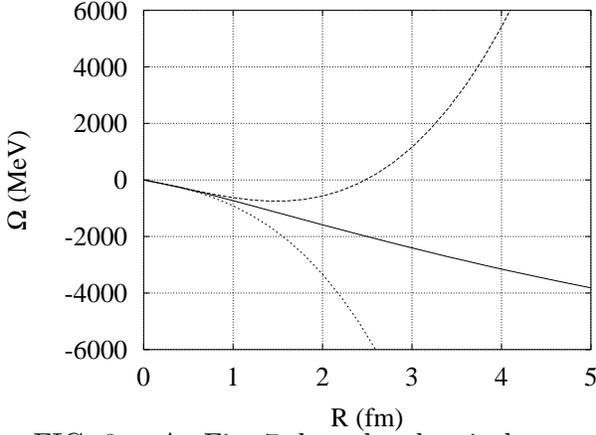}
\caption[]{
As Fig.~\ref{fig:o2vac100}, but the chemical potentials are now
$\mu_u=400$~MeV $\mu_s=100$~MeV, and the phase equilibrium
temperature corresponding to these values is $T_0=100.0$~MeV.
Again the chemical potentials are fixed, and the temperatures
are, top to bottom: $T=105,100,90$~MeV.
The minimum in the thermodynamic potential is quite deep
at these values of the parameters.}
\label{fig:o2vac400}
\end{figure}

We now investigate the effect of letting the $s$ quark have its
own chemical potential. To this end we show 
Figs.~\ref{fig:o2vac100} and \ref{fig:o2vac400}.
In Fig.~\ref{fig:o2vac100} the $s$ quark chemical potential
is 4 times greater than the light quark chemical potential,
whereas in Fig.~\ref{fig:o2vac400} it is converse.
The conclusion to be drawn from these figures is that 
a large $s$ quark chemical potential tends to wash out the minimum in the
thermodynamic potential, present when the light quark chemical 
potential is less than or equal to the $s$ quark chemical potential.
This behavior can be traced back to the positive surface tension
contribution from the massive $s$-quarks to the thermodynamic potential.

Finally, we underline that the interesting effect of bubble formation 
is due to the fact that the presence of a surface alters the 
density of states of particles.
Although these finite-size effects are 
negligible at large radii of the bubble or droplet,
they may have important implications 
for the phase transition as a whole.

\section{Conclusion}
\label{sec.concl}
We have seen that the thermodynamic potential of a hadron bubble
embedded in quark-gluon plasma, exhibits a minimum
at a radius $R \simeq 1$~fm,
even at temperatures somewhat above the bulk transition
temperature. Thus, within the model described here, 
a homogeneous plasma of size larger than a few fermi
is an impossibility.

Regarding the relevance of these results in connection with current and
forthcoming ultrarelativistic heavy ion collisions, we conclude that
{\em if} indeed a quark-gluon plasma is formed in the course of such a 
collision, then (according to the model considered here) this plasma phase
will contain bubbles of hadrons of radius $R \simeq 1$~fm.
{\em If} such hadronic bubbles form inside the plasma, the observable
effects are likely to include some blurring of the plasma signatures,
since this reduces the effective plasma volume.

We emphasize that these conclusions may well be 
model dependent, inasmuch the minimum of the thermodynamic
potential is of modest depth. 
However, the phenomenon of spontaneous creation 
of stable hadronic bubbles in a quark-gluon plasma, does seem to be
well established within the the model discussed here 
\cite{us,marsve,michael}.
The main question is now, whether these hadronic bubbles 
are of physical nature,  or merely an artifact of the model.
Certainly, it would improve confidence in these results 
if other models of QCD, and eventually lattice calculations, 
were to yield similar results.

\acknowledgments
JM was supported in part by the Theoretical Astrophysics Center
under the Danish National Research Foundation. We thank Michael
Christiansen for useful discussions.

\end{document}